\documentclass[preprint]{iucr}
     \journalcode{J}
\usepackage{amsmath}
\usepackage{url}
\usepackage[caption=false]{subfig}
\usepackage[colorinlistoftodos]{todonotes}
\begin{document}                  



\title{A Semi-Supervised Approach for Automatic Crystal Structure Classification}


\author[a,b]{Satvik}{Lolla}
\author[c]{Haotong}{Liang}
\author[b,c,d]{A. Gilad}{Kusne}
\author[c,e]{Ichiro}{Takeuchi}
\cauthor[b,c]{William}{Ratcliff}{william.ratcliff@nist.gov}

\aff[a]{Poolesville High School, Poolesville, MD 20837, \country{USA}}
\aff[b]{NIST Center for Neutron Research, NIST, Gaithersburg, MD, 20899, \country{USA}}
\aff[c]{Department of Materials Science and Engineering, University of Maryland, College Park, MD 20742, \country{USA}}
\aff[d]{Materials Measurement Laboratory, NIST, Gaithersburg, MD, 20899, \country{USA}}
\aff[e]{Maryland Quantum Materials Center, College Park, MD, 20742 \country{USA}}

\maketitle                        

\begin{synopsis}
Semi-supervised model to predict crystal structure from powder neutron diffraction patterns.
\end{synopsis}

\begin{abstract}
The structural solution problem can be a daunting and time consuming task. Especially in the presence of impurity phases, current methods such as indexing become more unstable. In this work, we apply the novel approach of semi-supervised learning towards the problem of identifying the Bravais lattice and the space group of inorganic crystals.  Our semi-supervised generative deep learning model can train on both labeled data- diffraction patterns with the associated crystal structure, and unlabeled data, diffraction patterns that lack this information. This approach allows our models to take advantage of the troves of unlabeled data that current supervised learning approaches cannot, which should result in models that can more accurately generalize to real data. In this work, we classify powder diffraction patterns into  all 14 Bravais lattices and 144 space groups (we limit the number due to sparse coverage in crystal structure databases), which covers more crystal classes than other studies. Our models also drastically outperform current deep learning approaches for both space group and Bravais Lattice classification\cite{billengePDF, garcia2019learning, ryu2019deep, ziletti2018insightful, tiong2020identification} using less training data.
\end{abstract}
\section{Introduction}

The first step to understanding the properties of a crystalline material at a microscopic level starts with identifying the crystal structure. However, identifying the crystal structure is nontrivial. The first part of crystal structure determination is indexing. There are several programs which can be used, such as Dicvol06\cite{dicvol}, Topas\cite{Topas}, GSAS II\cite{gsas} or N-TREOR\cite{treor,treor90}. These programs output a set of space groups and lattice parameters that could represent the crystal. Using Le Bail\cite{lebail} and Pawley\cite{pawley} refinements, the space group that fits the diffraction pattern the best can be identified. Rietveld \cite{rietveld1967line, rietveld1969profile} refinement can then be applied to profile the lattice parameters and check the space group. In the presence of impurity phases, this approach becomes more expensive as peaks must be selected manually or tolerance levels must be tuned to discard a certain number of peaks.\par
One of the approaches to identifying the position of atoms in a crystal is the charge flipping algorithm (CFA) \cite{CFA_first, Palatinus_summary, baerlocher2007charge}. CFA is an iterative approach that relies on Fast Fourier Transforms to determine the crystal structure of a material\cite{fft,superflip}. For CFA, the unit cell and Bravais Lattice have to be known--that is, we must have already have succeeded with some degree of indexing.  CFA also cannot handle impurity phases, which are prevalent in many real-world samples.\par
To automate this task, recent research looks to supervised neural networks (NN)s to analyze diffraction patterns. Supervised learning is an approach that seeks to learn a functional mapping between data and their labels. The benefit of NNs is that they, unlike CFA, do not require additional parameters such as the Bravais Lattice or lattice parameters. Although some approaches use NNs to aid in the Rietveld refinement\cite{ozaki2020automated, chang2020deep, reviewpaper, reviewpaper2}, others use NNs to classify diffraction patterns based on the crystal structure. These classifiers can be trained to identify impurity phases and can be tailored towards specific detectors or parameters. For example, Ryu et al.\cite{ryu2019deep} trained a NN to classify diffraction patterns of crystals that had defects. Liu et al.\cite{billengePDF} used the pair density function with powder neutron diffraction data for space group classification. Ziletti et al.\cite{ziletti2018insightful} used a convolutional neural network to classify simulated single crystal diffraction x-ray image data into 8 space groups. \par

A number of studies represent powder diffraction patterns as 2D images. However, the information is inherently one dimensional. Previous groups likely used the image approach to take easy advantage of trained models developed by the machine learning community. Unfortunately, this one to many mapping could introduce more complexities to the model. Garcia-Cardona et al.\cite{garcia2019learning} was one of the only studies to examine neutron scattering data and used a 1-dimensional (1D) approach with simulated powder diffraction data to both differentiate perovskites into 5 crystal systems and tune the lattice parameters using regression. This study only looked at a small subset of crystals \par
A significant challenge with NNs is that they struggle to generalize to new datasets. Most models that predict the space group of a material use less than 100 space groups in their training dataset, which limits their application to new diffraction patterns. However, large labeled diffraction datasets are often rare as labeling them is an expensive task. For this reason, we use a semi-supervised model, which takes advantage of both labeled and unlabeled data during training\cite{SGAN, zhu2009introduction,kingma2014semi,kipf2016semi}. We employ a generative network that can extract features from the unlabeled data distribution and match these features with the corresponding crystal structure. This allows semi-supervised learning to be used on more datasets, especially ones where labels are not available.\par
In this study, we propose a 1D semi-supervised model for Bravais Lattice and space group classification using powder neutron diffraction data. Our NNs are trained with data spanning 144 space groups, making our network more applicable than previous studies. The models used in this study are freely available and can be downloaded.\cite{github}
\section{Methods}
\subsection{Data}
\begin{figure}
  \includegraphics[width=6.5cm]{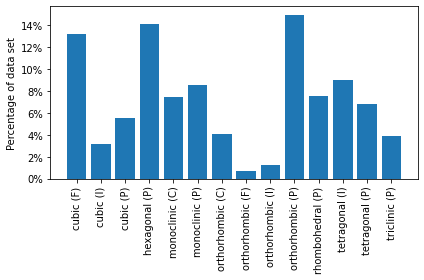}
  \includegraphics[width=6.5cm]{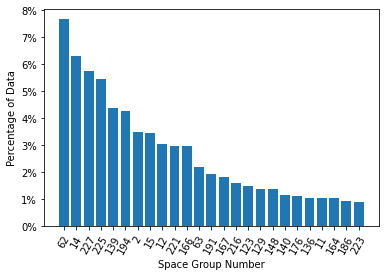}
  \par(a)\hspace{6.5cm}(b)
  \caption{Shows the (a) Bravais Lattice and (b) space group class distribution of the simulated data.}
  \label{data_imbalance}
\end{figure}

To test our approach under conditions where we know the correct answer, we work with simulated datasets. Our data was taken from the Inorganic Crystal Structure Database (ICSD), which contains structural information about more than 210,000 crystals \cite{ICSD, disclaimer}. A total of 138,362 diffraction patterns were simulated using Topas\cite{Topas}. For the Bravais Lattices, we combine the rhombohedral and tetragonal classes for a total of 14 classes with "F", "I", "P" and "C" representing the face centered, body centered, primitive and base centered lattices respectively. We note that there is an inherent class imbalance in the ICSD, as shown in Figure \ref{data_imbalance}. The most prevalent classes in this dataset were the primitive hexagonal, the face centered cubic, and the primitive orthorhombic lattices. The least represented lattices were the face centered orthorhombic and body centered orthorhombic lattices.\par
For the space groups, we use 136,454 of the simulated diffraction patterns. We used only the space groups that had more than 50 diffraction patterns, leaving us with 144 out of the 225 space groups present in the ICSD.
 The most frequent space group is 62 (Pnma) which is orthorhombic, and accounts for the disproportionately large orthorhombic (P) diffraction patterns in the ICSD dataset. A complete list of the 144 space groups used is shown in Table \ref{sgs}.\par
\begin{figure}
    \centering
      \includegraphics[width=6.5cm]{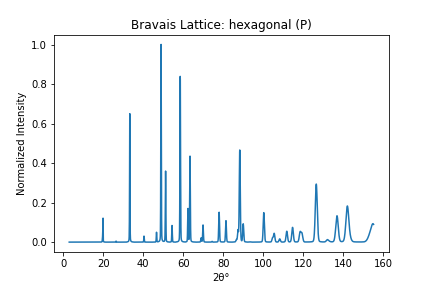}
      \includegraphics[width=6.5cm]{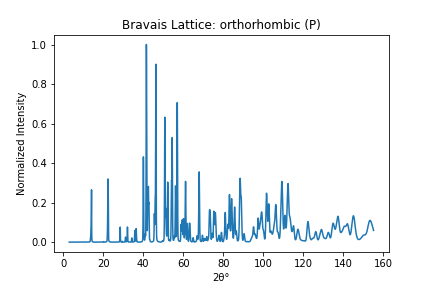}
      \par (a)\hspace{6.5cm}(b)
    \caption{Example 1D diffraction patterns for (a) Hexagonal (P) crystals and (b) Orthorhombic (P) crystals}
    \label{patterns}
\end{figure}

\begin{table}
    \centering
    \begin{tabular}{|c|c|}
    \hline
    \textbf{Crystal System}&\textbf{Space Groups}\\
    \hline
    Triclinic&1, 2\\
    \hline
    Monoclinic&4-15\\
    \hline
    Orthorhombic&18-20, 26, 29, 31, 33, 34,36, 38, 40-47, 51-53, 55-67, 69-74\\
    \hline
        Tetragonal&\begin{tabular}{@{}c@{}}82, 85-88, 92, 96, 99, 100, 107, 109, 113, 114, 119 \\ 121, 122, 123, 125-131, 135-137, 139-142\end{tabular}\\
    \hline
        Trigonal&143, 144, 146-148, 150, 152, 154-157, 159-167\\
    \hline
        Hexagonal&173-176, 180, 182, 185-187, 189-194\\
    \hline
        Cubic&197-201, 203-206, 212, 213, 215-218, 220, 221, 223-227, 229, 230\\
    \hline
    \end{tabular}
    \caption{List of space groups sorted by crystal system.}
    \label{sgs}
\end{table}
In this study, we use a one-dimensional approach rather than the traditional two-dimensional image one. Our dataset consists of diffraction patterns of powders. Examples of the 1D diffraction patterns are shown in Figure \ref{patterns}. To normalize these diffraction patterns, we divide all intensities in each diffraction pattern by the maximum intensity. This ensures that the new maximum intensity is equal to 1 and the minimum is equal to 0.
\subsection{Models}
We use two approaches to classify the diffraction patterns: a supervised approach using Convolutional Neural Networks (CNNs) and a semi-supervised approach using a Semi-Supervised Generative Adversarial Network (SGAN).
\subsection{Supervised Model}\label{resnet description}
We use a 1D ResNet-18, a Residual Network\cite{ResNet}, model to identify the crystal structure of the diffraction patterns. ResNets are examples of Convolutional Neural Networks (CNNs), which are commonly used for image classification network algorithms. CNNs consist of convolutional layers, which are responsible for extracting high-level features, such as edges and colors, from images. These layers are used to create a feature map consisting of the most relevant characteristics of the image. To create this map using 1D data, a filter of size $n$ is applied to a larger sequence with size $m$, and the dot product of every $n$ consecutive values and the filter is computed. This generates a smaller matrix that only includes the relevant features\cite{cnn}.\par
\begin{figure}
    \includegraphics[width=2cm]{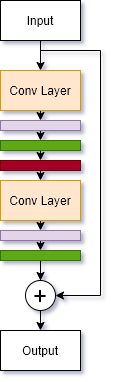}
    \includegraphics[width=5cm]{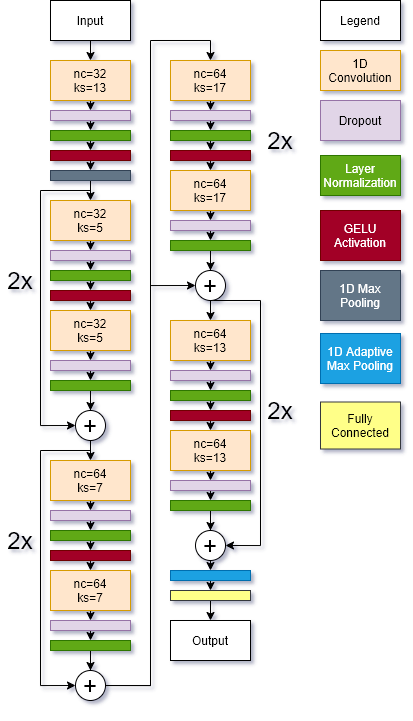}
    \par (a)\hspace{3.5cm}(b)\hspace{1.5cm}
    \caption{(a) shows ResNet block used in this study and (b) shows the ResNet-18 architecture. In both figures, the orange, purple, green, red, gray, blue, and yellow layers represent 1D Convolutional Layers, Dropout layers, Layer Normalization layers, GeLU activation functions, 1D Max Pooling layers, 1D Adaptive Max Pooling Layers, and Fully Connected layers respectively. The white circles with "+" signs in them represent the addition of two layers. In (b), each ResNet block was repeated twice, shown by the "2x" next to each block. The "nc" and "ks" represent the number of channels and the kernel stride for each convolutional layer.}
    \label{resnet_arch}
\end{figure}

A ResNet was used in this study to overcome the degradation problem, which occurs when neural networks are too dense that the accuracy saturates and then quickly degrades\cite{ResNet}. ResNets are characterized by their residual blocks, which contain convolutional layers with an identity function. Figure \ref{resnet_arch} shows the model architecture used for the ResNet-18, and includes an example of a ResNet block used in this model. During training, we randomly selected 90\% of the data to use as the training set and the remaining 10\% of the data was used to test the model. This testing dataset was distinct from the training one, so the model did not learn from the testing data.
\subsection{Semi-Supervised Model}
We also use a Semi-Supervised Generative Adversarial Network (SGAN)\cite{SGAN, GAN, improvedgans}. The SGAN consists of two models: a Generator and a Discriminator. The Generator tries to fool the Discriminator with fake diffraction patterns while the Discriminator aims to differentiate between real and fake diffraction patterns. The Discriminator also classifies the real labeled data into the corresponding crystal structure class. 
\subsubsection{Generator}
\begin{figure}
    \centering
    \includegraphics[width=6cm]{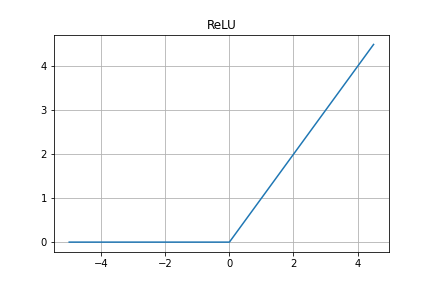}
    \includegraphics[width=6cm]{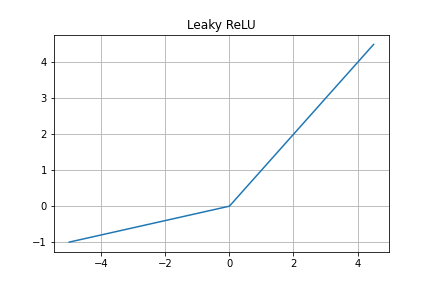}
    \par (a)\hspace{6cm}(b)\par
    \includegraphics[width=6cm]{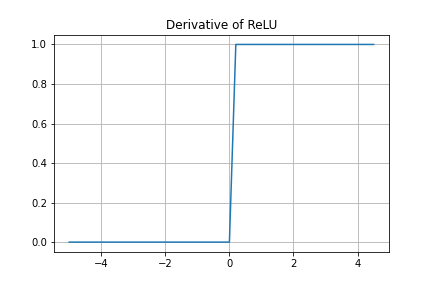}
    \includegraphics[width=6cm]{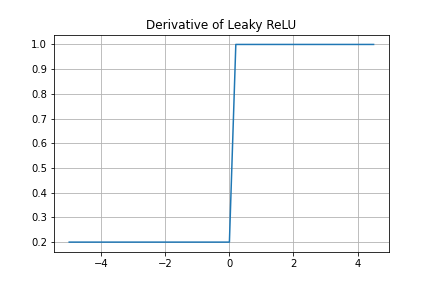}
    \par (c) \hspace{6cm}(d)\par
    
    \caption{Shows (a) ReLU, (b) Leaky ReLU with $\alpha=0.2$, (c) Derivative of ReLU, (d) Derivative of Leaky ReLU with $\alpha=0.2$}
    \label{relu}
\end{figure}

\begin{figure}
    \centering
    \includegraphics[height=7cm]{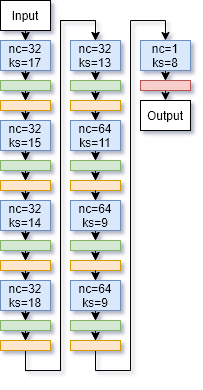}
    \caption{Shows the architecture for the Generator. The blue, greed, orange and red boxes represent 1D Convolutional Transpose Layers, LeakyReLU activations, Batch Normalization layers, and Sigmoid activations respectively. The "nc" and "ks" in the blue boxes represent the number of channels and the kernel size respectively for each 1D Convolutional Transpose Layer.}
    \label{g arch}
\end{figure}

The purpose of the Generator is to sample the latent space, a high dimensional feature space, to generate realistic diffraction patterns. The inputs to the Generator were sampled from a random normal distribution with a mean of 0 and standard deviation of 1. The Generator consists of 1D Convolutional Transpose layers, 1D Batch Normalization Layers, and a LeakyReLU activation function. The Convolutional Transpose layers are used to upsample the data\cite{dcgan, cnntranspose}. The Batch Normalization layers standardize the output of each layer, which reduces error when the model tries to generalize to new inputs\cite{batchnorm} and has also shown to reduce mode collapse, a major problem in GANs\cite{dcgan}. Mode collapse occurs when the generator only produces a few distinct diffraction patterns despite the latent space input. The Leaky ReLU\cite{leakyrelu} activation with $\alpha=0.2$ is used rather than ReLU to reduce the vanishing gradients problem\cite{dcgan}. Graphs of the ReLU and the Leaky ReLU activation functions are shown in Figure \ref{relu}. For negative values, the derivative for the Leaky ReLU function is equal to $\alpha$, but for the ReLU function, it is equal to 0. By having a non-zero derivative for all values, the Leaky ReLU is used to combat the sparse gradient problem that occurs while training GANs. Due to our normalization method, which was dividing all values in a diffraction pattern by the maximum intensity, the Discriminator's inputs were in the range from 0 to 1. For this reason, a Sigmoid activation was applied to the last layer of our Generator, rather than the hyperbolic tangent function recommended by Salisman et al.\cite{improvedgans} Figure \ref{g arch} shows the model architecture of the Generator.

\subsubsection{Discriminator}
\begin{figure}
    \centering
    \includegraphics[height=10cm]{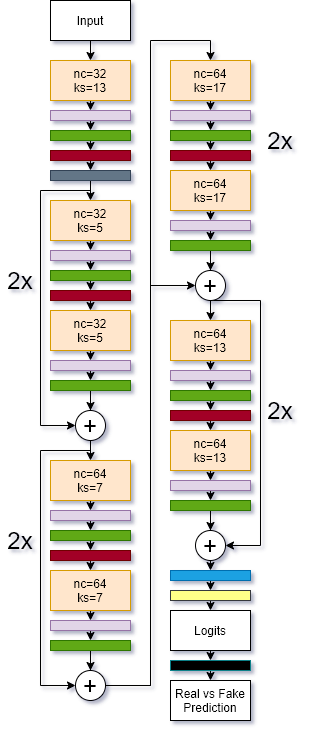}
    \caption{Model architecture for ResNet-18 discriminator. The orange, purple, green, red, gray, blue, and yellow layers represent 1D Convolutional Layers, Dropout layers, Layer Normalization layers, GeLU activation functions, 1D Max Pooling layers, 1D Adaptive Max Pooling Layers, and Fully Connected layers respectively. The white circles with "+" signs in them represent the addition of two layers. In (b), each ResNet block was repeated twice. The "nc" and "ks" represent the number of channels and the kernel stride for each convolutional layer.}
    \label{disc arch}
\end{figure}

The Discriminator has two objectives: differentiate between real and generated data, and classify the real data into the correct class. To do this, we use the same 1D ResNet-18 model described in Section \ref{resnet description}, but an activation function to the last fully connected layer, as proposed by Salisman et al\cite{improvedgans}. This activation function is shown in Equation \ref{act}, and is a version of the softmax activation. In this equation, $l_k(x)$ represents the logit for class $k$ with data $x$. By doing this, we eliminate the need for a second output layer, and instead use only the logits from the classification layer. By applying this activation function, diffraction patterns with larger logits, which signify more confident predictions, will be classified as "real" whereas diffraction patterns with smaller logits will be classified as "fake." This encourages the Discriminator to be more confident in its predictions, which sharpens the decision boundary between classes. The Discriminator's architecture is shown in Figure \ref{disc arch}.

\begin{equation}
\begin{split}
D(x)&=\frac{Z(x)}{Z(x)+1}\\
Z(x)&=\sum_{k=1}^K \exp(l_k(x))\\
\end{split}
\label{act}
\end{equation}

While training the Discriminator, there are two modes: supervised and unsupervised. During unsupervised training, the Discriminator acts the same way it would in a regular GAN as it tries to determine that the generated diffraction patterns are fake and the data drawn from the unsupervised set is real. In the supervised mode, the Discriminator is trained to predict the class label for real samples. Training in the unsupervised mode can help the Discriminator extract features from the data and training on the supervised data will allow the Discriminator to use those extracted features for classification.

\subsubsection {Loss Functions and Objective Functions}

The modified min-max loss proposed by Goodfellow et al.\cite{GAN} was used for the adversarial loss between the networks. The Generator's objective function that it tries to maximize is shown in Equation \ref{gen_obj}. $\theta_g$ represents the parameters in the Generator, $z^{(i)}$ represents the random values in the latent space. $G(z^{(i)})$ is the generated diffraction pattern from the Generator and $-\log (D(G(z^{(i)})))$ is the probability that the Discriminator predicts that the generated pattern is real.
\begin{equation}
\nabla \theta_g \frac{1}{m}\sum_{i=1}^{m} -\log (D(G(z^{(i)})))
\label{gen_obj}
\end{equation}

For the discriminator, the objective function to be maximized is shown in Equation \ref{disc_obj}. $\theta_d$ represents the parameters in the Discriminator, $x^{(i)}$ is unsupervised real data, and maximizing $D(x^{(i)})$ implies that the model can identify real data. Like the Generator's objective function, $z^{(i)}$ represents the random values in the latent space and $G(z^{(i)})$ is the generated diffraction pattern from the latent space. Increasing the value of $1-\log (D(G(z^{(i)})))$ shows that the Discriminator can determine that the generated patterns are fake. Equation \ref{disc_obj} also includes the categorical cross entropy loss, which is shown in the term $\sum_{i}^{C}t_i\log(s_i)$. Here, $C$ represents the number of classes, $t_i$ shows whether the $i$th class is the label of the diffraction pattern and $s_i$ is the discriminator's prediction.
\begin{equation}
\nabla \theta_d \frac{1}{m}\sum_{i=1}^{m} [\log (D(x^{(i)}))+ \log (1-D(G(z^{(i)})))] + \sum_{i}^{C}t_i\log(s_i)
\label{disc_obj}
\end{equation}

\subsubsection{Training Details}
\begin{figure}
    \centering
    \includegraphics[width=8cm]{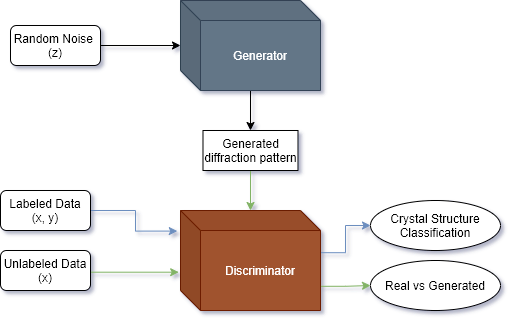}
    \caption{Shows the architecture for the Semi-Supervised GAN.}
    \label{sgan arch}
\end{figure}

Figure \ref{sgan arch} shows the training pipeline used in the SGAN. We train our SGAN using four different amounts of labeled training data. In all scenarios, we randomly select 10\% of the data as testing data, which is distinct from the labeled training data and the unlabeled training data. In the first scenario, we use 5\% of the data as labeled training data and 85\% as unlabeled training data. In the second, we use 10\% of the data as labeled training data and 80\% as unlabeled training data. In the third, we use 25\% of the data as labeled training data and 65\% of the data as unlabeled training data and finally we use 50\% of the data as labeled training data with 40\% of the data as unlabeled training data. We also train our supervised ResNet with the same 5\%, 10\%, 25\% and 50\% of the data to compare the accuracy of the SGAN to the purely supervised approach.

Table \ref{hyper} shows the hyperparameters used in the ResNet and the SGAN.
\begin{table}
    \centering
    \begin{tabular}{|c|c|c|c|}
    \hline
    \textbf{Hyperparameter}&\textbf{Supervised}&\textbf{Generator}&\textbf{Discriminator}\\
    \hline
    Optimizer&Adam&Adam&Adam\\
    \hline
    Learning Rate&1e-4,&1e-6& 5e-6\\
    \hline
    Dropout Rate&0.1& None& 0.1\\
    \hline
    Batch Size&64&32&32\\
    \hline
    Nonlinear Activations&GELU&Leaky ReLU\& Sigmoid&GELU\\
    \hline
    \end{tabular}
    \caption{Hyperparameters used in the supervised ResNet, Generator, and Discriminator}
    \label{hyper}
\end{table}
We used Pytorch\cite{torch} as a deep learning framework. To accelerate training, each model was trained on 8 NVIDIA Tesla V100 Tensor Cores.
\section{Results and Discussion}
\subsection{Supervised Model}
\begin{figure}
    \centering
    \includegraphics[width=14.3cm]{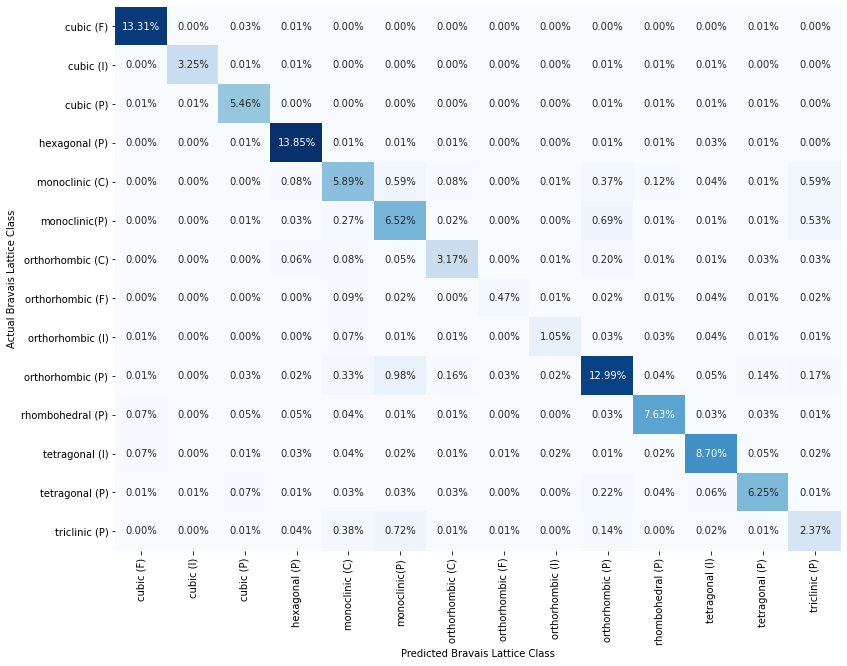}
    \caption{Shows the confusion matrix for the supervised ResNet-18}
    \label{Supervised_conf}
\end{figure}

Our supervised ResNet trained on 90\% of the dataset had an accuracy of 88\%. The confusion matrix for the Bravais Lattice model is shown in Figure \ref{Supervised_conf}. By plotting the predicted Bravais Lattice against the actual Bravais Lattice, the confusion matrix provides more information about the sets of classes that the network misclassified. If the model had a perfect testing accuracy, the values along the principal diagonal would sum to 100\% as the network would have classified every diffraction pattern correctly. Again, there is some clear imbalance in the sampled ICSD dataset, with orthorhombic (F) and orthorhombic (I) having the least samples. From the confusion matrix, we can see that despite the fact that orthorhombic (P) is the most prevalent class, the model misclassifies some of these as monoclinic (P) crystals. The network also has trouble differentiating between triclinic (P) and monoclinic (P) diffraction patterns as both of these classes have low symmetries, agreeing with previous studies\cite{garcia2019learning}.\par
For the space group identification, our model had a top 1 accuracy of 80.6\% and a top-5 accuracy of 90.27\% across all 144 space groups. Accuracy was measured by dividing the number of correctly classified diffraction patterns in the testing set by the total number of patterns in the testing set. Top-5 accuracy is the percentage of samples for which the actual space group was one of the model's top 5 predictions. This outperforms most current models that we are aware of: Liu et al. \cite{billengePDF} used machine learning with a pair-wise distribution function with a top-1 accuracy of 71\% and a top-5 accuracy of 90\% across 45 space groups. Tiong et al.\cite{tiong2020identification} classified x-ray diffraction data into 8, 20, 49, and 72 space groups. Their accuracy decreased from 99\% to 80\% for 8 and 72 space groups respectively, implying that this accuracy would decrease further if their model was trained on more space groups. Aguilar et al.\cite{sg_1d} had a top-2 accuracy greater than 80\% across all space groups but used a dataset consisting of 650,000 diffraction patterns, more than 5 times the size of the dataset used in this study. However, Aguilar et al. used a 1D network, suggesting that a 1D approach can lead to more accurate predictions.  We note that we did not take advantage of data-augmentation.
\begin{table}
    \centering
    \begin{tabular}{|c|c|c|c|}
    \hline
    \textbf{Study}&\begin{tabular}{@{}c@{}}\textbf{Number of}\\ \textbf{Space Groups}\end{tabular}&\textbf{Type of Data}&\textbf{Accuracy}\\
    \hline
    Our Model&144&Powder Diffraction&80\%\\
    \hline
    Liu et al.\cite{billengePDF}&45&PDF&71\%\\
    \hline
    Ziletti et al.\cite{ziletti2018insightful}&8&2D XRD&96.3\%\\
    \hline
    Tiong et al.\cite{tiong2020identification}&72&2D XRD&80.2\%\\
    \hline
    \end{tabular}
    \caption{Comparison to other supervised space group classifiers}
    \label{sg_comp}
\end{table}
\subsection{Semi-Supervised Model}
\begin{figure}
    \centering
    \includegraphics[width=10cm]{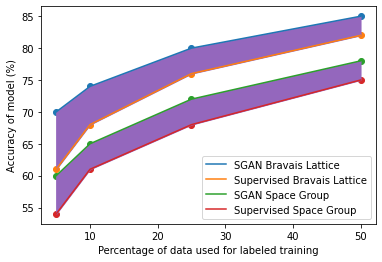}
    \caption{Graph comparing the accuracies of the SGAN and supervised ResNet for Bravais Lattice and Space Group identification.}
    \label{accCompFig}
\end{figure}

We compare the accuracy of the SGAN to the accuracy of the supervised model in Table \ref{acc_comp}. The SGAN consistently outperforms the purely supervised model, showing that the semi-supervised approach has the potential to be more applicable in the real world. A graph comparing the accuracy of the supervised and semi supervised models is shown in Figure \ref{accCompFig}. This graph shows that although the accuracy of the SGAN is impacted by a lack of data, the difference between the SGAN's accuracy and the supervised model's accuracy is greatest when only 5\% of the data is used.
\begin{table}
    \centering
    \begin{tabular}{|c|c|c|c|c|}
    \hline
    Percentage of data&
    \begin{tabular}{@{}c@{}}Supervised \\ Bravais Lattice\end{tabular}
    &\begin{tabular}{@{}c@{}}SGAN \\ Bravais Lattice\end{tabular}&
    \begin{tabular}{@{}c@{}}Supervised \\ Space Group\end{tabular}&
    \begin{tabular}{@{}c@{}}SGAN \\ Space Group\end{tabular}\\
    \hline
    5\%&61\%&70\%&54\%&60\%\\
    10\%&68\%&74\%&61\%&65\%\\
    25\%&76\%&80\%&68\%&72\%\\
    50\%&82\%&85\%&75\%&78\%\\
    \hline
    \end{tabular}
    \caption{Comparing the accuracy of the SGAN to the purely supervised approach with different amounts of labeled training data for both Bravais Lattice and Space Group classification}
    \label{acc_comp}
\end{table}
\section{Conclusion}
In this study, we use both CNNs and a semi-supervised GAN to investigate supervised and semi-supervised approaches for crystal structure classification. We demonstrate that semi-supervised GANs can prove to be more accurate with limited amounts of labeled data for Bravais Lattice and space group classification. Further, we explore a 1D approach rather than a traditional 2D one. Our 1D model is more accurate than 2D image models, which agrees with previous results in literature. Our semi-supervised model is also more applicable to real datasets which will lack large amounts of labeled data.\par
In the future, we would like to train the SGAN to identify impurity phases and to test the method on real data sets.
\ack{Acknowledgements}\par
The authors are grateful to Austin McDannald and Hui Wu from NIST and Brian Toby from Argonne National Laboratory. We acknowledge Stephan R\"{u}hl for the ICSD. We acknowledge support from the Center for High Resolution Neutron Scattering (CHRNS) a national user facility jointly funded by the NCNR and the NSF under Agreement No. DMR-2010792. The work at the University of Maryland was supported by NIST grant No. 60NANB19D027.\par

\referencelist{}

\end{document}